\documentclass[twoside,twocolumn,9pt]{article}
\usepackage{extsizes}
\usepackage[super,sort&compress,comma]{natbib} 
\usepackage[version=3]{mhchem}
\usepackage[left=1.5cm, right=1.5cm, top=1.785cm, bottom=2.0cm]{geometry}
\usepackage{balance}
\usepackage{mathptmx}
\usepackage{sectsty}
\usepackage{graphicx} 
\usepackage{lastpage}
\usepackage[format=plain,justification=justified,singlelinecheck=false,font={stretch=1.125,small,sf},labelfont=bf,labelsep=space]{caption}
\usepackage{float}
\usepackage{fancyhdr}
\usepackage{fnpos}
\usepackage[english]{babel}
\addto{\captionsenglish}{%
  
}
\usepackage{array}
\usepackage{textcomp}
\usepackage{droidsans}
\usepackage{charter}
\usepackage[T1]{fontenc}
\usepackage[usenames,dvipsnames]{xcolor}
\usepackage{setspace}
\usepackage[compact]{titlesec}
\usepackage{hyperref}

\usepackage{epstopdf}

\definecolor{cream}{RGB}{222,217,201}

\begin{document}

\pagestyle{fancy}
\thispagestyle{plain}
\fancypagestyle{plain}{
\renewcommand{\headrulewidth}{0pt}
}

\makeFNbottom
\makeatletter
\renewcommand\LARGE{\@setfontsize\LARGE{15pt}{17}}
\renewcommand\Large{\@setfontsize\Large{12pt}{14}}
\renewcommand\large{\@setfontsize\large{10pt}{12}}
\renewcommand\footnotesize{\@setfontsize\footnotesize{7pt}{10}}
\makeatother

\renewcommand{\thefootnote}{\fnsymbol{footnote}}
\renewcommand\footnoterule{\vspace*{1pt}%
\color{cream}\hrule width 3.5in height 0.4pt \color{black}\vspace*{5pt}} 
\setcounter{secnumdepth}{5}

\makeatletter 
\renewcommand\@biblabel[1]{#1}            
\renewcommand\@makefntext[1]%
{\noindent\makebox[0pt][r]{\@thefnmark\,}#1}
\makeatother 
\renewcommand{\figurename}{\small{Fig.}~}
\sectionfont{\sffamily\Large}
\subsectionfont{\normalsize}
\subsubsectionfont{\bf}
\setstretch{1.125} 
\setlength{\skip\footins}{0.8cm}
\setlength{\footnotesep}{0.25cm}
\setlength{\jot}{10pt}
\titlespacing*{\section}{0pt}{4pt}{4pt}
\titlespacing*{\subsection}{0pt}{15pt}{1pt}

\fancyfoot{}
\fancyfoot[LO,RE]{\vspace{-7.1pt}\includegraphics[height=9pt]{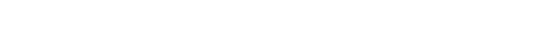}}
\fancyfoot[CO]{\vspace{-7.1pt}\hspace{13.2cm}\includegraphics{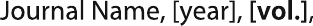}}
\fancyfoot[CE]{\vspace{-7.2pt}\hspace{-14.2cm}\includegraphics{head_foot/RF}}
\fancyfoot[RO]{\footnotesize{\sffamily{1--\pageref{LastPage} ~\textbar  \hspace{2pt}\thepage}}}
\fancyfoot[LE]{\footnotesize{\sffamily{\thepage~\textbar\hspace{3.45cm} 1--\pageref{LastPage}}}}
\fancyhead{}
\renewcommand{\headrulewidth}{0pt} 
\renewcommand{\footrulewidth}{0pt}
\setlength{\arrayrulewidth}{1pt}
\setlength{\columnsep}{6.5mm}
\setlength\bibsep{1pt}

\makeatletter 
\newlength{\figrulesep} 
\setlength{\figrulesep}{0.5\textfloatsep} 

\newcommand{\topfigrule}{\vspace*{-1pt}%
\noindent{\color{cream}\rule[-\figrulesep]{\columnwidth}{1.5pt}} }

\newcommand{\botfigrule}{\vspace*{-2pt}%
\noindent{\color{cream}\rule[\figrulesep]{\columnwidth}{1.5pt}} }

\newcommand{\dblfigrule}{\vspace*{-1pt}%
\noindent{\color{cream}\rule[-\figrulesep]{\textwidth}{1.5pt}} }

\makeatother

\twocolumn[
  \begin{@twocolumnfalse}
{\includegraphics[height=30pt]{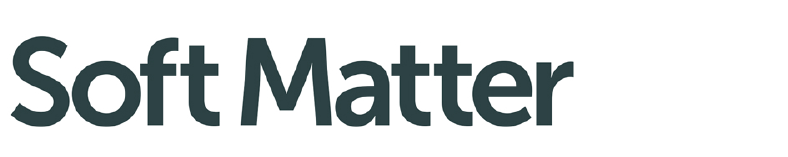}\hfill\raisebox{0pt}[0pt][0pt]{\includegraphics[height=55pt]{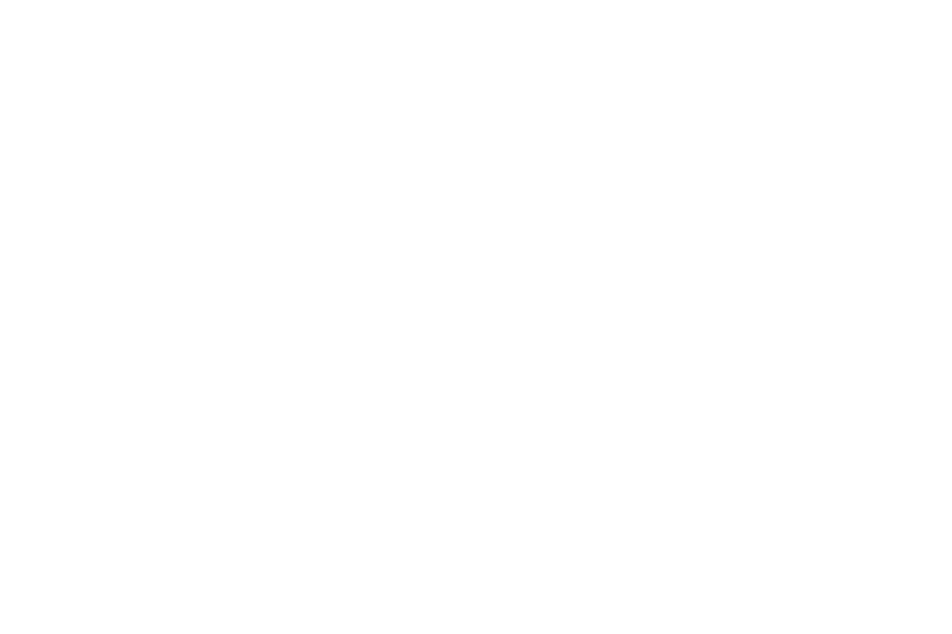}}\\[1ex]
\includegraphics[width=18.5cm]{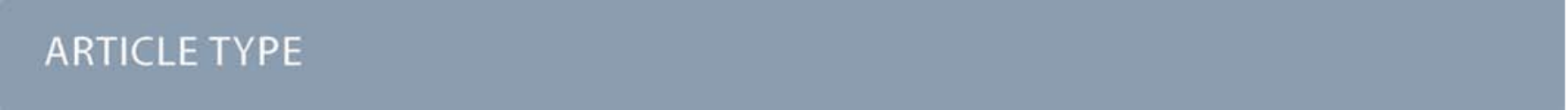}}\par
\vspace{1em}
\sffamily
\begin{tabular}{m{4.5cm} p{13.5cm} }

\includegraphics{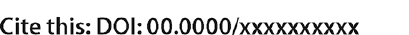} & \noindent\LARGE{\textbf{3D motion of flexible ferromagnetic filaments under rotating magnetic field $^\dag$}} \\
\vspace{0.3cm} & \vspace{0.3cm} \\

 & \noindent\large{Abdelqader Zaben,\textit{$^{a}$} Guntars Kitenbergs\textit{$^{a}$} and Andrejs C\={e}bers\textit{$^\ast$}{$^{a}$}} \\

\includegraphics{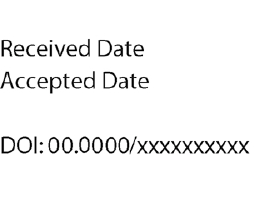} & \noindent\normalsize{Ferromagnetic filaments in a rotating magnetic field are studied both numerically and experimentally. The filaments are made from micron-sized ferromagnetic particles linked with DNA strands. It is found that at low frequencies of the rotating field a filament rotates synchronously with the field and beyond a critical frequency it undergoes a transition to a three dimensional regime. In this regime the tips of the filament rotate synchronously with the field on circular trajectories in the plane parallel to the plane of the rotating field. The characteristics of this motion found numerically match the experimental data and allow us to obtain the physical properties of such filaments. We also discuss the differences in behaviour between magnetic rods and filaments and the applicability of filaments in mixing.} 
\end{tabular}

 \end{@twocolumnfalse} \vspace{0.6cm}

 ]

\renewcommand*\rmdefault{bch}\normalfont\upshape
\rmfamily
\section*{}
\vspace{-1cm}


\footnotetext{\textit{$^{a}$~MMML lab, University of Latvia, Jelgavas 3, Riga, LV-1004, Latvia; E-mail: andrejs.cebers@lu.lv}}

\footnotetext{\dag~Electronic Supplementary Information (ESI) available: Supplementary Videos and their description. See DOI: 10.1039/cXsm00000x/}




\section{Introduction}

Magnetic filaments are the subject of growing interest due to their applications, for example, as mixers for microfluidics \cite{10} or in creating self-propelling microdevices capable of carrying cargo for targeted transport \cite{20}. Different methods of driving magnetic particle systems were recently reviewed by \citet{30}. Magnetic rods under the action of a rotating field are of special interest for use in microrheological measurements by magnetic rotational spectroscopy \cite{40,50}. The behavior of ferromagnetic rods under the action of a rotating magnetic field is interesting. It is known that for frequencies larger than the critical, a rigid ferromagnetic rod has two different regimes - one is an infinite in-plane rotation with a periodic angular velocity, and the other is a precessional motion around an axis inclined with respect to the plane of the rotating field. This situation is structurally unstable and at some small perturbation, for example, due to the Earth's magnetic field, only the precessional regime survives. For a rigid ferromagnetic rod this was recently illustrated experimentally \cite{60}. Previously we have numerically studied the planar regime of a flexible ferromagnetic rod under the action of a rotating magnetic field \cite{4}. It was shown that a filament has synchronous or asynchronous regimes of rotation depending on the frequency of the rotating field.

In the present paper the stability of the asynchronous planar regime of the flexible ferromagnetic rod is studied both experimentally and numerically. Using ferromagnetic filaments whose physical properties were measured previously \cite{80}, we found that the planar asynchronous regime is unstable at frequencies above the critical and the precessional regime is established. In this regime the tips of the filament move in the antiphase along circular trajectories in the planes parallel to the plane of rotating field. We discuss the differences between ferromagnetic and paramagnetic filament motion, as well their applicability to mixing.

\section{Experimental}
\label{sec:exp}
The filament synthesis was done as presented in details by \citet{K.thesis}. The filaments are made by linking  $4.26$~$\mu$m ferromagnetic particles (Spherotech, $1\%$w/v) with $1000$~bp long biotinized DNA strands (ASLA biotech,$192~\mu$g/ml). The chromium oxide coated polystyrene particles are functionalized with streptavidin; which bonds with the biotin in the DNA ends and forms filaments. The samples are made by mixing $0.5$~ml of $10\%$ TE buffer solution ($\eta = 0.01$~P, pH = 7.5) with $10$~$\mu$l of DNA giving a resulting concentration of $6.2$~nM. $2$~$\mu$l of particle solution is then added ($0.004\%~w/v$), and the sample is placed between two Neodymium magnets ($\approx 500$~Oe) for two minutes to allow particle alignment. Chains are formed with length $L$ ranging from $8~\mu$m to $80~\mu$m.  

$20$~$\mu$l of the sample is placed in a fluidic cell for observation. The cell is made by spacing two glass sides with $211$~$\mu$m thick adhesive tape. The fluidic cell is then placed under an optical microscope (Leica DMI3000B) using a $40x$ objective, bright field configuration. The magnetic field is generated using a coil system powered by Kepco BOP 20-10M power supplies, providing a field strength up to $120$~Oe, with a maximum frequency of $50$~Hz. The field profile is generated by a LabVIEW code through National Instruments data acquisition card (NI PCI-6229), which is connected to the power supplies. The images are acquired using Basler ac1920-155um camera with a fixed frame rate setting (up to $100$~fps).

Following the above procedure, the filaments were found to be randomly distributed with various lengths across the cell. A rotating field profile with a frequency $f=1$~Hz and a field strength $H=17.2$~Oe is first applied for 15 minutes to allow free particles to connect to the filaments. Due to the density difference, the filaments tend to sediment at the bottom of the cell. The filaments were found to have a slight movement during rotation, probably because of the particle to wall interaction. 

\begin{figure}[!h]
\centering
  \includegraphics[width=\columnwidth]{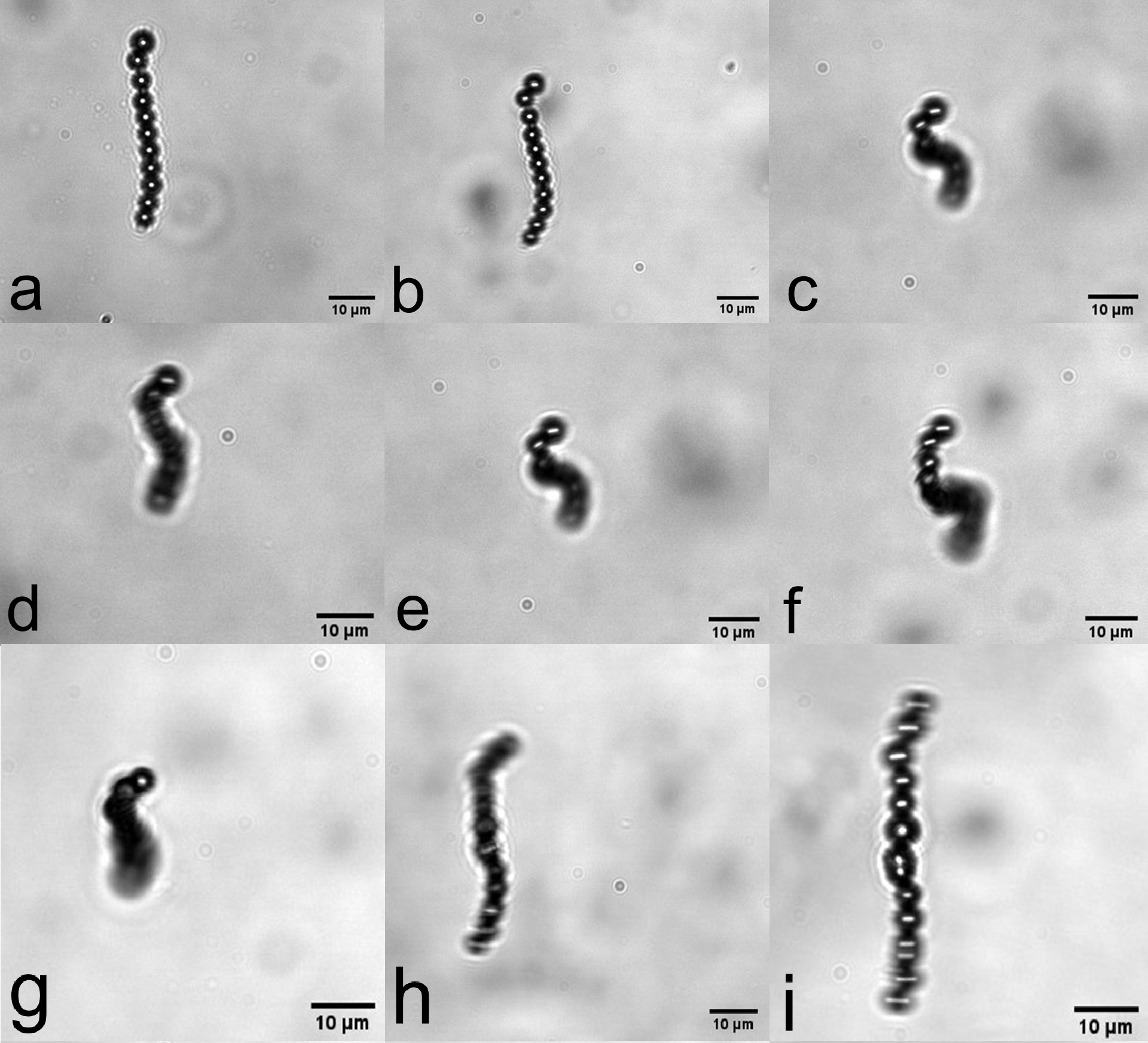}
  \caption{Shapes of deformed filaments under rotating magnetic field. Filament with $L=41$~$\mu$m at $H=17.2$~Oe for frequencies (a) $f=1$~Hz, (b) $f=4$~Hz, (c) $f=9$~Hz. Filaments at $H=17.2$~Oe and $f=8$~Hz for filaments with lengths (d) $L=33$~$\mu$m, (e) $L=41$~$\mu$m, (f) $L=65$~$\mu$m. Filament with $L=46$~$\mu$m at $f=5$~Hz and magnetic fields (g) $H=4.3$~Oe, (h) $H=12.9$~Oe, (i) $H=25.8$~Oe.}
  \label{fig:1}
\end{figure} 

The experiments were done by applying a rotating field to samples of filaments with different lengths $L$, at multiple field strengths $H$ and frequencies $f$. Some examples of deformed filament shapes under a rotating field in $x-y$ plane are shown in Fig.~\ref{fig:1}. The effect of increasing the frequency $f$ can be seen in Fig.~\ref{fig:1}(a)-(c), where length $L$ and field strength $H$ are constant. At low frequencies, the filament rotates synchronously with the magnetic field having an orientation lag, which results in small deformations as visible in Fig.~\ref{fig:1}(a) and (b) and supplementary Video S1. We define the critical frequency ($f_c$) as the smallest frequency at which the steady state of the deformed filament has moved out of the plane of the rotating field. Experimentally, this can be seen as blurry parts of the filament, which have moved out of the focus plane. At higher frequencies ($f$ > $f_c$), the filament tip motion was found to have a circular path (data not shown). An example of a deformed filament shape can be seen in Fig.~\ref{fig:1}(c) and supplementary Video S2. Fig.\ref{fig:1}(d) - (f) show the 3D shape for filaments of different lengths $L$, operating at a constant frequency ($f > f_c$) and field strength $H$. The effect of increasing the field strength $H$ is shown in Fig.~\ref{fig:1}(g) - (i), for which the filament length $L$ and frequency $f$ are constant. 

\begin{figure}[!h]
\centering
  \includegraphics[width=\columnwidth]{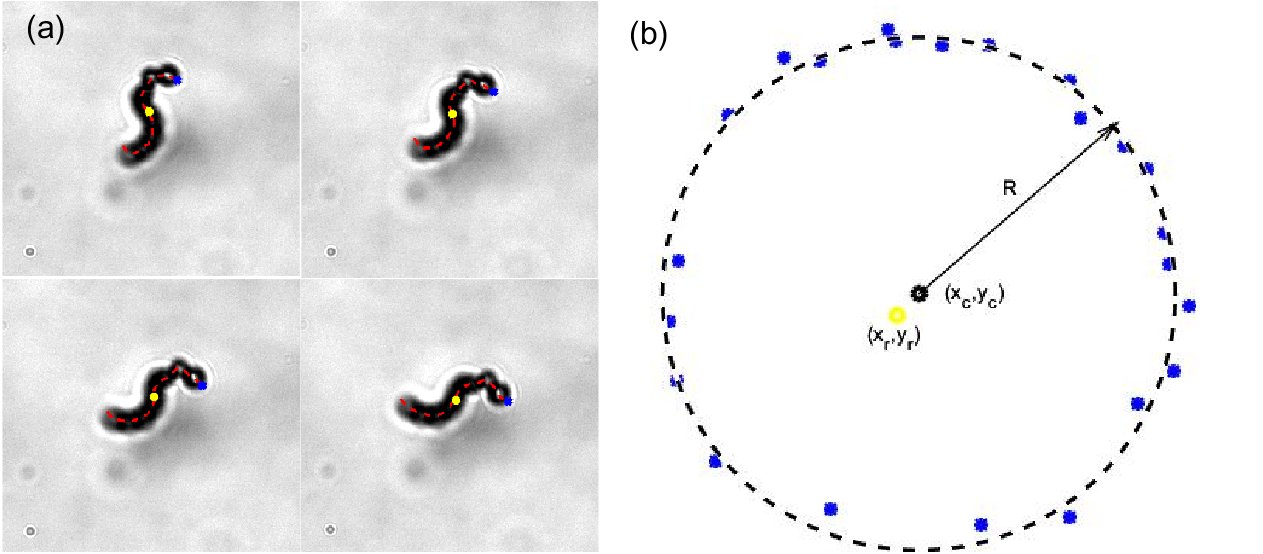}
  \caption{An illustration of how radius $R$ is for each filament. (a) Four consecutive images of a filament, which is rotating in 3D. Red lines indicate the centre line. The blue points shows the tracked, while yellow points represent the tracked centre, which is used for reference. (b) The tracked tip coordinates (blue points) are fitted with a circle (dashed line). Yellow point ($(x_r,y_r)$) is the reference and black point ($(x_c,y_c)$) denotes the fitted centre.}
  \label{fig:2}
\end{figure}
Image processing was done using Matlab. First, the images were segmented based on filament intensity. Second, the centre line of the filament was obtained using (NRBC) package \cite{river}. To characterize the filament deformation we introduce the radius $R$. An illustration of radius $R$ determination can be seen in Fig.~\ref{fig:2}. Then the coordinates of one of the filament tips $(x_t,y_t)$ and centre $(x_i,y_i)$ obtained from the detected centre line are recorded for each frame, shown as the blue and yellow points in Fig.~\ref{fig:2}(a) respectively. As the filaments are slightly moving during experiments, we use the filament centre coordinates $(x_r,y_r)$, yellow point shown in Fig.~\ref{fig:2}(b), in the first image as a reference point. Then the difference between the centre coordinates $(x_i,y_i)$ and $(x_r,y_r)$ is calculated to obtain the new tip coordinates after translation. The tip coordinates, shown as scattered blue points in Fig.~\ref{fig:2}(b), are fitted with a circle equation. This gives the radius $R$ and the centre of rotation $(x_c,y_c)$ (black point). The differences between the blue points and the fitted circle allows us to define an experimental error as $\sigma$. It is important to also mention that the rotation of the filament is not symmetric with respect to the centre of the filament. To account for this, the length $L$ is taken as $2\cdot R_0$, where $R_0$ is the radius of an undeformed filament, which rotates at a low frequency. This is the reason why only one tip of each filament is tracked.

\begin{figure}[!h]
\centering
  \includegraphics[width=\columnwidth]{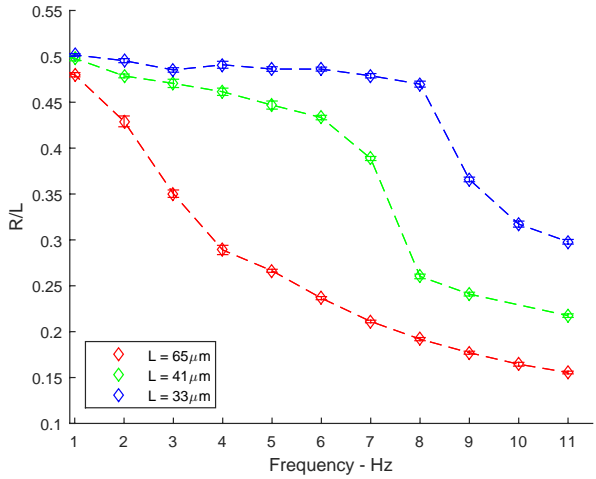}
  \caption{Experimental data for the relationship between R/L and frequency; filaments with different length and operating at field strength H =$17.2$~Oe.}
  \label{fig:3}
\end{figure}

Following the image processing procedure mentioned above, we find the radius $R$ at each frequency $f$ and magnetic field $H$. In order to make results comparable, we scale radius $R$ with length $L$. This means that for an undeformed filament $R/L=0.5$. Fig.~\ref{fig:3} shows the $R/L$ dependence on frequency $f$ for three filaments with different length $L$ at the same magnetic field $H=17.2$~Oe. For frequencies below critical $f_c$, the results show that the filaments have minor deformations reaching R/L $\sim$ 0.45. This is followed by a rapid drop of $R/L$ for frequencies above critical. A further increase in the frequency was found to have smaller influence on $R/L$ drop. It is clearly visible that a longer filament has a lower critical frequency than shorter filaments, if the field is kept constant.

\section{Model and numerical results}

The ferromagnetic filament is described by a modified Kirchhoff model in which the action of magnetic torques is taken into account \cite{1,2}. The force $\vec{F}$ in the cross section reads
\begin{equation}
\vec{F}=-A_{b}\frac{\partial^3 \vec{r}}{\partial l^{3}}+\Lambda\frac{\partial \vec{r}}{\partial l}-M\vec{H}
\label{Eq:10}
\end{equation}
where $l$ is the natural  parameter of the centerline, $A_{b}$ is the bending modulus of the rod, the linear density of the magnetization $M$ is along the local tangent direction and the Lagrange multiplier $\Lambda$ enforces the inextensibility of the filament. The dynamics of the filament are described by the partial differential equation
\begin{equation}
\zeta\frac{\partial \vec{r}}{\partial t}=-A_{b}\frac{\partial^{4} \vec{r}}{\partial l^{4}}+\frac{\partial}{\partial l}\Bigl(\Lambda\frac{\partial \vec{r}}{\partial l}\Bigr)
\label{Eq:11}
\end{equation}
and the term with the magnetic field enters only through the boundary conditions. These in the present case are taken as for the rod with free and unclamped ends: $\vec{F}|_{0,L}=0;~\partial^{2}\vec{r}/\partial l^{2}|_{0,L}=0$ ($L$ is the length of the rod). The Eq.\ref{Eq:11} is solved for each time step by an implicit scheme and the inextensibility is accounted for by the projection operator $P=I-J^{T}(J\zeta^{-1}J^{t})^{-1}J$, where $J=\partial g_{i}/\partial \vec{r}_{k}$ is the $p\times3(p+1)$ matrix for the rod with $p+1$ marker points and $p$ constraints $g_{i}=(\vec{r}_{i+1}-\vec{r}_{i})^{2}=h^{2}$. 

We introduce the length $L$ as the characteristic length scale and the elastic relaxation time $\tau_{e}=\zeta L^{4}/A_{b}$ as the time scale, where $\zeta=4\pi\eta$ and $\eta$ is the viscosity of the medium. As a result the dynamics of the filament are determined by two dimensionless parameters - the magnetoelastic number $Cm=MHL^{2}/A_{b}$ and $\omega\tau_{e}$.

Previously, this algorithm was used to simulate the buckling of a ferromagnetic rod at field inversion \cite{1}, anomalous orientation in an AC field of high frequency \cite{2}, self-propulsion \cite{3} and other phenomena. The behavior of the ferromagnetic rod in a rotating field was also considered \cite{4}. Here we give evidence that the planar asynchronous regime is unstable and the filament relaxes to a three-dimensional precessional regime around the vector of the angular velocity of the rotating field.

\begin{figure}[h]
\centering
 \includegraphics[width=\columnwidth]{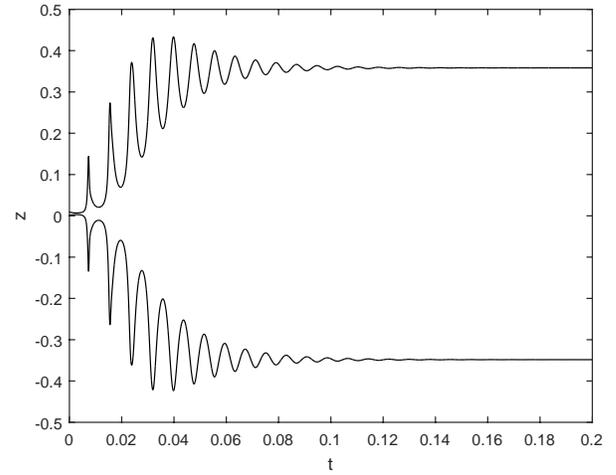}
  \caption{Time dependence of filament tips' $z$ coordinates show formation of a precessional regime. $Cm=50$, $\omega\tau_{e}=1000$.}
 \label{fig4}
\end{figure}

\begin{figure}[h]
\centering
 \includegraphics[width=\columnwidth]{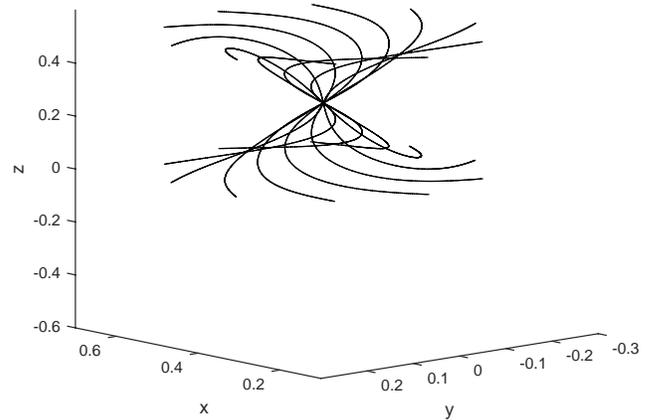}
  \caption{Configurations of the filament in precessional regime. $Cm=220$, $\omega\tau_{e}=4500$.}
 \label{fig5}
\end{figure}

The algorithm is run from initial conditions until a dynamic equilibrium is reached. In Fig.~\ref{fig4} the instability of the planar regime is illustrated by the time dependence of the $z$ coordinates of the filament tips at $Cm=50$ and $\omega\tau_{e}=1000$. This is caused by a tiny deviation from the rotational  plane - at $t=0$ the initial coordinate of one tip is set to $z=0.004$. When the precessional regime is established, the tips move in anti-phase along circular trajectories. Let us note that in this regime the motion of the filament is synchronous with the rotating field. The numerically calculated filament configurations in the precessional regime for $Cm=220$ and $\omega\tau_{e}=4500$ are shown in Fig.~\ref{fig5}. Video examples of planar and 3D regimes can be seen in supplementary Videos S3 and S4. 

\begin{figure}[!h]
\centering
 \includegraphics[width=\columnwidth]{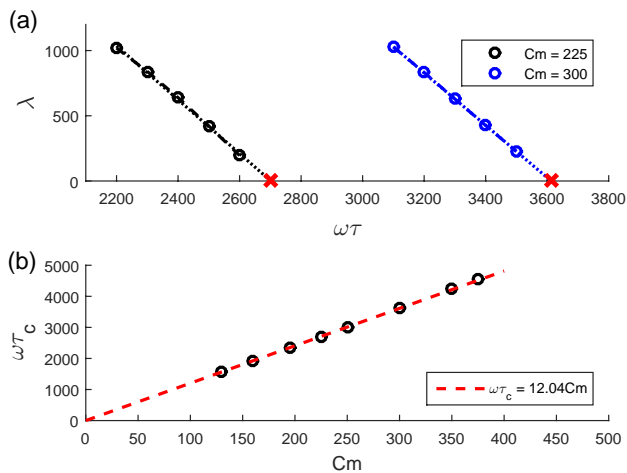}
  \caption{Critical frequency $(\omega\tau_{e})_{c}$ is proportional to the magnetoelastic number $Cm$. (a) An example of finding critical frequencies by decrement ($\lambda$) analysis for $Cm=225$ and $Cm=300$. Circles represent numerical data, lines - linear fits and red $\times$ correspond to critical frequencies $(\omega\tau_{e})_{c}$. (b) Critical frequency as a function of $Cm$. Circles are numerical data and the dashed line is the linear fit.}
 \label{fig6}
\end{figure}

The transition to the precessional regime occurs when the frequency is above the critical for the synchronous regime. Its numerical value may be found by the study of the relaxation of the filament tip to the plane of the rotating field which for small perturbations occurs according to the law $z=z(0)\exp{(-\lambda t)}$. For each $Cm$ we find the decrements $\lambda$ as a function of frequency $\omega\tau_{e}$, as is shown in Fig.~\ref{fig6}(a). By extrapolating data to $\lambda=0$, we find critical frequencies ${\omega\tau_{e}}_{c}$ for multiple magnetoelastic numbers $Cm$. These values are shown in Fig.~\ref{fig6}(b). The data are well described by a linear dependence, which is ${\omega\tau_{e}}_{c}=12.04Cm$. The value of the slope is very close to $12$, which was predicted for a ferromagnetic rod \cite{4}.

\begin{figure}[!h]
\centering
  \includegraphics[width=\columnwidth]{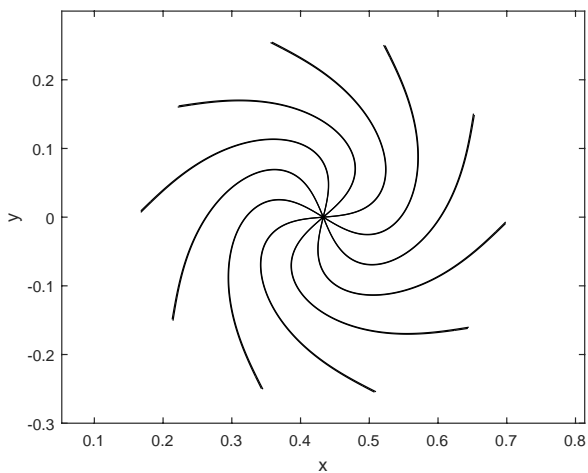}
  \caption{Configurations of the filament in precessional regime in the projection on the plane of rotating field ($Cm=220~\omega\tau_{e}=4500$). }
  \label{fig8}
\end{figure}

It should be remarked that the configurations of the filament in the precessional regime are quite strongly bent as may be seen from the experimental data Fig.~\ref{fig:1}. This is in qualitative agreement with the numerical simulation data, as shown in Fig.~\ref{fig8}., where the filament configurations for $Cm=220$ and $\omega\tau_{e}=4500$ are shown in the projection on $x,y$ plane.

\begin{figure}[!h]
\centering
  \includegraphics[width=\columnwidth]{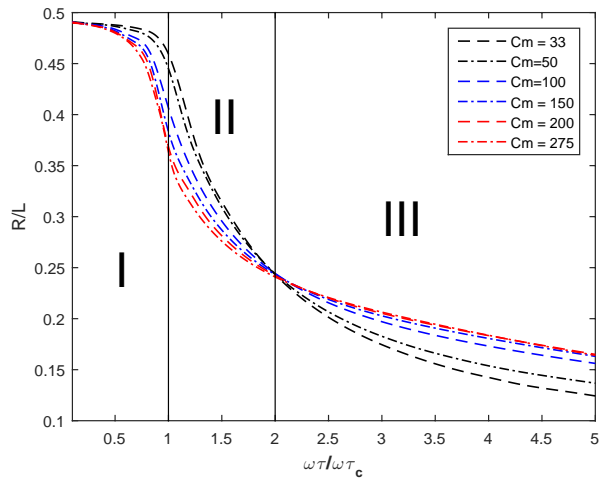}
  \caption{Numerical simulation results for filament deformations. Normalized deformation $R/L$ as a function of normalized frequency $\omega\tau/\omega\tau_c$.}
  \label{fig9}
\end{figure}

The linear dependence of the critical frequency allows us to qualitatively analyze the numerical results and divide the filament behaviour into three regimes. Also, for numerical simulations we can define a deformation measure as the radius $R$. Here it can be simply taken as the half distance between the filament tips in the $x-y$ plane. The corresponding relationship of $R/L$ as a function of frequency $\omega\tau$, normalised by $\omega\tau_c$, is shown in Fig.~\ref{fig9}. Region I: In-plane rotation with filament deformation depending on $Cm$. At critical frequency $\omega\tau_c$ filament at lower $Cm$ deforms less. The critical frequency corresponds to $\omega\tau/\omega\tau_c=1$. Region II: 3D rotations with a rapid drop in $R/L$. Region III: A further increase in $\omega\tau/\omega\tau_c$ has a lower rate of $R/L$ drop. It should be noted that operating at higher $Cm$ ($Cm > 200$) has no influence on $R/L$ at region III. 

\section{Comparison of experimental and numerical \\results}

Previously experimental and numerical results were presented separately. In this section we compare them both qualitatively and quantitatively.

\begin{figure}[!h]
\centering
 \includegraphics[width=\columnwidth]{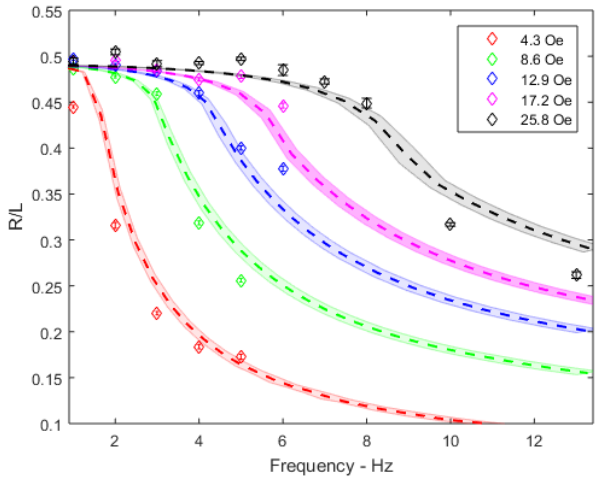}
  \caption{Filament deformation comparison between experiment and numerical results via $R/L$ dependence on frequency. The colored diamonds denote experimental data for filament with $L = 46$~$\mu$m operating at different field strength $H$. The dashed lines represents numerical simulations scaled with values of parameters obtained from the fit for different $Cm$ values as shown in Table.\ref{table:1}. The shaded area represents the error margins obtained from fitting.}
 \label{fig11}
\end{figure}

Using $R/L$ allows us to directly compare the filament deformations. Data in Fig.~\ref{fig:3} and Fig.~\ref{fig9} show a very similar behavior for the deformation dependence on frequency. There is an initial phase with little deformations, followed by a sharp drop and finishing with a much slower decrease. To perform a quantitative comparison, we made an experiment with the same filament ($L=\text{const}$) for five different magnetic fields $H$ and multiple rotating field frequencies $f$ for each of them. The experimental data are visible in Fig.~\ref{fig11} as diamonds with different colours.

To fit the numerical data to the experimental results, we introduce two parameters $A$ and $B$. The magnetic field $H$ is linked to the magnetoelastic number $Cm$ by $A=Cm/H$. The frequency $f$ is linked to the dimensionless frequency $\omega\tau_e$ by $B={\omega\tau_{e}/f}$.

\begin{table}[h!]
\centering
\begin{tabular}{|c | c|   } 

 \hline
 $H$, Oe & $Cm$  \\
  \hline
 $4.3$ & $31$   \\ 
  \hline
 $8.6$ & $62.5$  \\
  \hline
  $12.9$ & $93.8$ \\
  \hline
  $17.2$ & $125$ \\
 \hline
  $25.8$ & $187.5$  \\ 
 \hline
\end{tabular}
\caption{Experimental field strength $H$ and corresponding magnetoelastic numbers $Cm$, as found by non-linear fit.}
\label{table:1}
\end{table}

$A$ and $B$ are found with Matlab by a nonlinear least squares fit of all experimental data simultaneously. For that we take a 2D interpolation (scatteredInterpolant) of many numerical curves $R/L=func(\omega\tau_e,Cm)$ and build a custom function. Experimental errors are used as weights. Taking care of local minima, this results in $(A\pm\Delta A)=(125\pm 2.2)~\text{Oe}^{-1}$ and $(B\pm\Delta B)=(241.6\pm 4.4)$~s. The fit allows us to relate values for each field strength $H$ with a magnetoelastic number $Cm$, which are shown in Table~\ref{table:1}.

With numerical results and parameters $A$ and $B$ it is possible to construct the best fits. These are shown in Fig.~\ref{fig11} as dashed lines of the same colour as the corresponding magnetic field. For a more obvious comparison we have added the possible fit errors with corresponding colour shades. They are constructed from the maximal positive and negative deviations, according to the errors. A reasonably good agreement is visible in the figure. For some points the fitted values have been slightly overestimated, while for others - underestimated. 

It is important to verify the feasibility of the fit parameters. As $\omega=2\pi f$, we can find the elastic relaxation time $\tau_e=B/2\pi=38.5$~s. Using the parameter definitions and the same approach as previously \cite{80}, we can find the physical properties of the filament, namely, magnetization per unit length $M=4.9\cdot10^{-7}$~emu and bending modulus $A_b=1.5\cdot10^{-12}$ erg$\cdot$cm. These values are close to the estimates we have found earlier \cite{80}, confirming the quality of the fit.

\begin{figure}[!h]
\centering
 \includegraphics[width=\columnwidth]{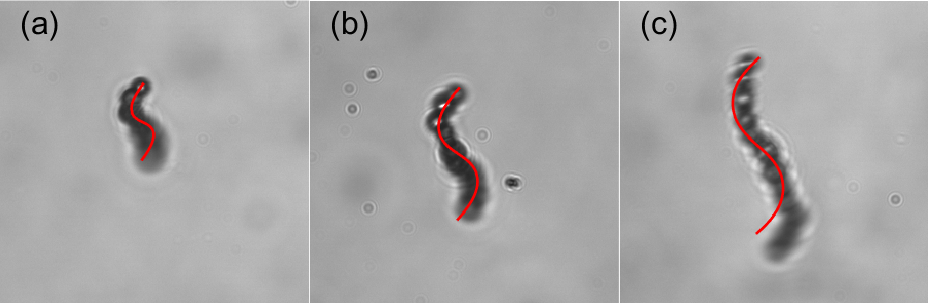}
  \caption{Direct comparison of filaments in experimental images and numerical simulations, using best fit, shows a qualitative agreement. Filament of $L=46$~$\mu$m at $f=5Hz$ and different $H$ is compared to numerical results at $\omega\tau=1208$ and different $Cm$. (a) $H=4.3$~Oe, $Cm=31$ , (b) $H=8.6$~Oe, $Cm=62.5$, (c) $H=12.9$~Oe, $Cm=93.8$.}
 \label{fig12}
\end{figure}

A direct way of comparing experiments with the numerical data is by looking at the filament shapes. The experimental system with an optical microscope allows us only to view the 2d shape of the filament in the $x-y$ plane, while the third dimension can be seen by the out of focus blur. Using the parameters $A$ and $B$, as well as knowing the filament length $L=46$~$\mu$m allows us to overlay experimental images with projections of numerical results on the $x-y$ plane. This can be seen in Fig.~\ref{fig12} for 3 different magnetic fields. A qualitative agreement is easily visible. Notable discrepancies can be seen for the lower part of the filament. This clearly justifies the chosen approximations in image processing, as described in \S\ref{sec:exp}., namely, using the centre of rotation for determining $R$, not the whole length. The behaviour of the chosen part of the filament is predictable.

\section{Discussion}

Our results clearly show that a ferromagnetic filaments in a rotating field can have motion in the third dimension.
However, this only happens above a critical frequency.
An out of plane motion - precession - was predicted for soft magnetic particles in a rotating field \cite{Cimurs}.
Later this was observed experimentally for ferromagnetic nanorods \cite{Kornev3D} and recently elaborated in detail both experimentally and theoretically by \citet{60}, where they show how a ferromagnetic rod beyond a critical frequency can have two possible regimes: unstable in-plane asynchronous back and forth motion and stable synchronous precession.
It is worth noting that the critical frequency $f_c$ both for ferromagnetic rods and filaments is proportional to the magnetic field $H$ and inversely proportional to the length squared $L^2$.

Qualitatively, the appearance of the 3D regime can be explained with the interplay of the magnetic torque, the filament elasticity, the hydrodynamic drag and a small perturbation.
For low frequencies the increasing frequency leads to higher drag, which is compensated by elastic deformations or reduced by the increase of the magnetic field and the corresponding torque \cite{80}. 
Above a critical frequency, the magnetic torque is insufficient to deform the filament more and the drag forces the filament to lose synchronization with the field, while remaining in the plane of the rotating field.
Surprisingly, a small perturbation, which might come from the Earth's magnetic field, is sufficient to rotate the deformed filament out of plane.
Coming out of the plane reduces the radius $R$, which results in reduced drag.
Torque is once again large enough to cause a synchronous rotation, but now for a filament that is deformed in 3D.
At least the appearance of stable precessing states for ferromagnetic rods is explained by analyzing the effect of small perturbations on the phase portrait \cite{60}.

Numerically, we find that the planar regime of the filament loses its stability at $\omega\tau_{e}=12\cdot Cm$ and the three dimensional precessional regime is established.
Thus, at the critical frequency $f_{c}$ the following relation is valid $2\pi f_{c}\zeta L^{2}/(12T/L)=1$ where $T$ is torque and $T/L=MH$ is the torque per unit length of the filament. 
It should be noted that taking $\zeta=4\pi\eta/\log{(L/d)}$ and $T/L=\pi^2\chi^{2}H^{2}d^{2}/6$, a similar relation is valid for the critical frequency of a chain of paramagnetic particles with a susceptibility $\chi$ and a diameter $d$ \cite{100}.
While relations for critical frequencies are similar, the important difference is in their behaviour beyond $f_c$.
A paramagnetic filament undergoes a planar back and forth regime, while a ferromagnetic filament transitions to the 3D regime.
This is in agreement with the behaviour of ferromagnetic and paramagnetic rods in a rotating field.
Beyond the critical frequency a paramagnetic rod has only the back and forth regime, while the ferromagnetic rods have both in-plane and out of plane behaviour, as discussed earlier.

For paramagnetic filaments at high frequencies of the rotating field a peculiar behaviour of filament coiling can be observed \cite{100}.
This behaviour was not seen for ferromagnetic filaments. It is probably due to the finite magnetic relaxation time of paramagnetic beads \cite{110}.

Another interesting point is the value of the bending modulus of the DNA linked ferromagnetic particle chain.
Here we find $A_{b}=1.5\cdot 10^{-12}$ erg$\cdot$cm. 
This is one order of magnitude larger than from the dipolar interactions $M^{2}/2$ ($M=4.9\cdot 10^{-7}$ emu) \cite{120}.
Therefore, the elastic deformations and the corresponding bending modulus in our experiments comes from the deformations of DNA linkers.
It is worth adding that the bending modulus of the DNA linker based filaments depends on their length.
As was shown with paramagnetic bead based filaments, the change in linker length can influence the resulting bending modulus across three orders of magnitude \cite{130}.

Rotating chains and filaments have a potential for mixing applications in microfluidics \cite{mixreview}. 
Many examples use paramagnetic particle chains, however, such filaments have a clear limitation at the critical frequency, above which filaments experience back and forth motion, which forces a reduction in mixing efficiency \cite{10}.
In comparison, ferromagnetic filaments have a transition to the three dimensional regime and continue to rotate with the angular frequency of the field.
This difference should enable efficient mixing also for higher frequencies and will be studied in detail in the near future.

\section{Conclusions}

In this study we have investigated the regimes that flexible ferromagnetic filaments undergo with the application of a 2D rotating magnetic field. It was found that for frequencies below the critical the filament rotates synchronously with the magnetic field with an orientation lag and a slightly deformed shape. For frequencies above the critical, the filament undergoes a transition to a three dimensional precessional regime, which is similar to the behaviour of ferromagentic rods. The critical frequency $f_c$ is lower for longer filaments and higher for stronger magnetic fields. The experimentally obtained characteristics of this regime are in good agreement with the results of numerical simulations.

\section*{Conflicts of interest}

There are no conflicts to declare

\section*{Acknowledgements}
A.Z. acknowledges support from the European Union\textquotesingle s Horizon 2020 research and innovation programme under grant agreement MAMI No. 766007, A.C. from M.era-net project FMF No.1.1.1.5./ERANET/18/04 and G.K. from PostDocLatvia grant No. 1.1.1.2/VIAA/1/16/197. Authors thank J. Cimurs for discussions.

\balance

\bibliography{rsc} 
\bibliographystyle{rsc} 

\end{document}